*[English Translation - Original Greek version follows]*

# Artificial Intelligence in Secondary Education: Educational Affordances and Constraints of ChatGPT-4o Use

Tryfon Sivenas[1], Panagiota Maragkaki[2]

*[1] Postdoctoral Researcher, National and Kapodistrian University of Athens, sivenastrif@primedu.uoa.gr*
*[2] Researcher, University of Nicosia, giwmar@icloud.com*

## ABSTRACT

The purpose of this study was to examine, from the perspective of secondary education students, the educational affordances and constraints of using Artificial Intelligence (AI) in teaching and learning. The sample consisted of 45 students from the 2nd year of General Lyceum (11th grade, ages 16–17), who, after becoming familiarized with ChatGPT-4o and completing six activities, filled in an open-ended questionnaire related to the research purpose. Open, axial, and selective coding of the data revealed that students recognize five educational affordances: the creation of new knowledge building on prior knowledge, immediate feedback, friendly interaction through messaging, ease and speed of access to information, and skills development. Concurrently, three main constraints were identified: content reliability, anxiety about AI use, and privacy concerns. The study concludes that students are positive toward AI use in education.

**Keywords:** Artificial Intelligence, Secondary Education, Educational Technology.

## INTRODUCTION

In recent years, rapid technological developments in Artificial Intelligence (AI) have attracted researchers' interest regarding its application and use across a number of fields (Xu et al., 2021). Education constitutes one such field (Holmes & Tuomi, 2022). Recent systematic reviews (Deng & Yu, 2023; Pérez et al., 2020) have highlighted that AI can contribute positively to reshaping the educational landscape (Holmes & Tuomi, 2022). To date, research on the application of AI in education has focused both on teachers (Zawacki-Richter et al., 2019) and on students (Deng & Yu, 2023). In a recent systematic review, Zhang and Tur (2024) report that AI facilitates adaptive learning, improves and enriches the learning environment, while simultaneously contributing to the development of a number of student skills such as active participation, initiative-taking, and creativity.

From recent studies (e.g., Adeshola & Adepoju, 2024), reviews (e.g., Zhai et al., 2021), and systematic reviews (e.g., Deng & Yu, 2023; Zhang & Tur, 2024), it emerges that the tool predominantly used in educational research on AI was ChatGPT-3.5. The Generative Pretrained Transformer (GPT) by OpenAI is a language model designed to produce text resembling human language and constitutes a subject of particular interest for educational use (Deng & Yu, 2023). ChatGPT is now in version 4o, which became available on May 13, 2024 (OpenAI, 2025), and offers without additional charge an increased number of features and capabilities. A characteristic example is the ability to create multimedia content (e.g., images via DALL-E 2.0) and content summarization of attached files, which in previous versions required a monthly subscription. Nevertheless, a small number of studies have utilized the new version, as confirmed by recent systematic reviews (e.g., Zhang & Tur, 2024).

Specifically, as Baig and Yadegaridehkordi (2024) note, most studies appear to focus predominantly on the use of ChatGPT-3.5, while fewer are those that use version 4.0 and its new capabilities. Additionally, the review of the research literature confirmed that most studies focus on higher education (Baig & Yadegaridehkordi, 2024), while the use of GPT-4o in secondary education has not been extensively examined. A further gap concerns the lack of studies examining the affordances that GPT-4o offers in education.

The purpose of this study is to address this research gap and examine the educational affordances and constraints of GPT-4o use with a sample of 45 students from the 2nd year of General Lyceum. Specifically, the research answers the following research questions:

1) What are the affordances of GPT-4o for teaching and learning?
2) What are the constraints that may limit the use of GPT-4o in teaching and learning?
3) Would secondary education students choose to use GPT-4o in teaching and learning?

## DEFINITION OF AFFORDANCES

The term affordance refers to the description of the offered capabilities and advantages that a technology can provide in education (Sivenas & Koutromanos, 2022a). Given that the use of GPT-4o in education may have wide application in the future, it is important to examine its affordances for teaching and learning. As Masoudi et al. (2019) note, examining the affordances of a technology in education helps in better understanding its characteristics and how they can contribute to teaching and learning. The term affordance was introduced by psychologist Gibson in the late 1970s (Gibson, 1977), while Norman (1999) elaborated on perceived affordances within the framework of Human-Computer Interaction (HCI).

Specifically, perceived affordances refer to what an individual perceives about an object, which will determine the way they use it (Norman, 1999). Norman (1999) states that "the term affordances focuses on the perceived and actual properties of an object, emphasizing the fundamental properties that determine exactly how an object can be used." Following the above, this study adopts Norman's approach, as it relates to the needs of this research and the interaction between humans and computers.

## REVIEW OF THE RESEARCH LITERATURE

A review of the research literature was conducted in the following databases: (1) IEEE Xplore, (2) ERIC, (3) SpringerLink, (4) Scopus, and (5) Science Direct. The following keywords were used: "affordances OR educational affordances OR constraints AND artificial intelligence, OR AI, OR chatbot, OR Chatgpt OR openAI."

The criteria were threefold and concerned (1) the search for studies from 2004 onwards, (2) studies in English and Greek, and (3) studies with empirical data.

The results of the literature review did not reveal studies examining the educational affordances and constraints of AI use in teaching and learning. On the other hand, it revealed a number of studies examining the educational affordances of corresponding emerging technologies.

One of the earliest studies is that of Conole and Dyke (2004), who examined the affordances of ICT and the degree to which they facilitate educational use. The researchers identified seven educational affordances, including access to a large volume of information, immediacy in rapidly changing information, and communication and collaboration. Similar studies on educational affordances were conducted by Dalgarno and Lee (2010), who examined the educational affordances of three-dimensional virtual learning environments; Song (2011), who examined the educational affordances of handheld PDAs; Bower and Sturman (2015), who examined the educational affordances of smart wearable technologies; and more recently, Sivenas and Koutromanos (2022b), who examined the educational affordances of drones for teaching and learning.

## METHODOLOGY

This study was conducted in May 2024 in three classes of the 2nd year of General Lyceum (11th grade, ages 16–17) in Attica, Greece. Specifically, the intervention took place in the school's computer laboratory, so that students had access to computers.

### *Participants*

Forty-five students from the 2nd year of General Lyceum participated in the study, of whom 26 (57.78%) were male and 19 (42.22%) were female. Of the 45 participants, 18 reported having used an AI application in the past.

### *Questionnaire*

Data collection was carried out through an anonymous online questionnaire (Google Forms) consisting of two parts. The first part included demographic questions (e.g., gender, age) and a question asking students whether they had previously used an AI application. The second part comprised five open-ended questions asking students about: (1) the educational affordances they identified in GPT-4o, (2) whether they would use GPT-4o for teaching and learning in the future, (3) how they envisioned using GPT-4o for teaching and learning, (4) the extent to which they would use GPT-4o for personal use, and (5) the constraints that could potentially inhibit AI use in teaching and learning. The above questions were adapted from similar studies in the research literature (e.g., Bower & Sturman, 2015; Sivenas & Koutromanos, 2022a).

### *Equipment*

A total of 20 desktop computers from the school's informatics laboratory were utilized. For the intervention, the ChatGPT desktop application was not installed locally; instead, students used the pre-installed browser to visit the OpenAI website.

### *Procedure*

The intervention was carried out in two stages. In the first stage, a 45-minute introductory presentation was given regarding AI, its historical background, its evolution, and its everyday applications. In the second stage, teachers guided students in creating individual GPT-4o accounts and familiarizing themselves with its environment. Subsequently, students were asked to complete six activities aimed at familiarizing them with GPT-4o's capabilities.

The first activity asked students to search for information from GPT-4o about a topic of interest. The second activity involved creating a personalized resume. Specifically, the teachers presented the following scenario: "Imagine that this summer when schools close, you want to take a summer job for July and August. Ask ChatGPT to create a resume based on your hobbies and the field you want to work in this summer." Students then provided personal details (e.g., name, surname, hobbies, language certifications, contact details) and GPT-4o created a personalized resume that students saved as a Microsoft Word document. The third activity involved using DALL-E 2.0 for image creation.

The fourth activity involved content summarization. Students were asked to search the internet for texts or documents related to a topic of interest, then upload the text or document to GPT-4o and request a content summary, followed by questions about the uploaded content. For the fifth activity, students were asked to use GPT-4o to create a quiz of five questions on a school subject they found difficult. Subsequently, students asked GPT-4o to answer the questions it had created. This activity aimed to demonstrate to students that GPT-4o can answer incorrectly. The sixth and final activity asked students to use GPT-4o to request information and explanation of difficult terms, with GPT-4o addressing different age groups. Specifically, teachers asked students to select a subject they found difficult (e.g., mathematics), then to choose a specific topic within that subject (e.g., quadratic equation). Students were then asked to type in GPT-4o: "Explain X to me as if you were addressing a 7, 10, and 14-year-old child." The duration of the second stage was 45 minutes.

Following this, students had 15 minutes of free use of GPT-4o. Finally, the online questionnaire was completed, which took fifteen minutes.

### *Data Analysis*

Data analysis was conducted by two specialists in ICT in education and was based on the principles of qualitative research and thematic analysis by Corbin and Strauss (2014). Specifically, open, axial, and selective coding of the data was performed (Corbin & Strauss, 2014). A similar methodological approach has been followed by previous studies in the research literature examining the educational affordances of technologies (e.g., Bower & Sturman, 2015; Koutromanos et al., 2024; Sivenas & Koutromanos, 2022b).

## RESULTS

The results revealed five educational affordances of AI, as shown in Table 1. Overall, the majority of participants (N = 42) recognized the benefits AI can offer to education. Below follows the analysis of each educational affordance.

### *Creating New Knowledge by Leveraging Prior Knowledge*

The majority of participants (N = 33) recognized the capability of acquiring new knowledge as a particularly important educational affordance. Specifically, students emphasized that they would use GPT-4o to discuss topics they already knew about or topics about which they had partial knowledge. This was because they understood during the intervention that GPT-4o can make errors. Conversely, a number of students (N = 13) reported that they would use GPT-4o for creating new knowledge without considering the possibility of error.

### *Immediate Content Feedback*

An educational affordance highlighted by a large proportion of students (N = 28) concerns the immediate feedback offered by GPT-4o. Specifically, two response categories were identified. The first concerns that immediate feedback relates to content written by the students themselves in GPT-4o that is, the conversation stage. Thus, they reported that GPT-4o helped them improve their text or describe more precisely what they were trying to write. The second response category concerns summarization—the intervention where students uploaded an attached file and requested content summarization. A characteristic example is the response of Student No. 14, who stated: "I uploaded a PDF with a book for my sister's master's that I had on my flash drive. The book is 500 pages, but with AI it did the summary in two minutes, it's very fast, I don't know if I would have had the patience to wait an hour for it though."

### *Friendly/Familiar Interaction Environment Through Messaging*

Students referred to the friendly user environment through messaging (N = 25). As they reported, this mode of communication is very familiar and friendly, being the communication method they choose for chatting with friends. This resulted in the conversation with GPT-4o feeling more familiar to them.

### *Ease and Speed of Receiving Information*

Another affordance highlighted by students concerns the speed with which they receive responses from GPT-4o (N = 18). Specifically, response time is particularly important for students, with several commenting that in case of slow response, they might not use it.

### *Development of a Number of Skills*

Several students indirectly referred to GPT-4o's contribution to developing skills (N = 11). More specifically, students reported they believed GPT-4o could help them improve in certain subjects, resulting in active participation in teaching. Other student examples focused on understanding the informatics course and specifically programming commands. The remaining skills mentioned included digital literacy (N = 8), computational thinking (N = 6), problem-solving (N = 4), critical thinking (N = 3), and communication (N = 2).

Regarding the final research question examining the constraints that may affect GPT-4o use in teaching and learning, three categories were identified (see Table 1), analyzed below.

### *Uncertainty Regarding the Truthfulness and Quality of Content*

The first constraint highlighted by several students concerned uncertainty regarding the truthfulness and quality of content (N = 16). Specifically, several students referred to the fact that after the intervention, they would not choose GPT-4o to acquire new knowledge, but would only request additional information on knowledge they already possessed. Conversely, some students noted that it acts as an inhibiting factor, since after each information search on GPT-4o, they would need additional time to search the internet to verify whether the results/data provided by GPT-4o were accurate.

### *Anxiety Regarding AI Use*

A number of students reported that GPT-4o use creates anxiety and stress (N = 8). Most responses were based on the content feedback capability. Most student comments concerned the fact that the text they wrote/created would never satisfy GPT-4o. A characteristic example is Student No. 32, who stated: "It stresses me that every time I ask how I can improve the text I wrote, it keeps making comments. Is my text never perfect? It keeps suggesting changes and how it can be improved; at some point it must stop and say it's OK." A smaller number of students referred to feeling anxious because, as Student No. 9 characteristically states: "...I can talk to an entity that knows everything and that is stressful."

### *Privacy Constraints*

Few students referred to GPT-4o's privacy constraints (N = 5). Most responses concerned students' concerns about whether someone could see the information they ask GPT-4o, as well as where that information is stored. Other responses concerned the protection of their privacy, where students did not wish to provide their name when registering on the OpenAI website.

**Table 1**

*Educational Affordances and Constraints of Artificial Intelligence Use in Secondary Education*

| Affordances of GPT-4o for teaching and learning | Constraints |
|---|---|
| Creating new knowledge by leveraging prior knowledge (N = 33) | Uncertainty regarding the truthfulness and quality of content (N = 16) |
| Immediate content feedback (N = 28) | Anxiety regarding AI use (N = 8) |
| Friendly/familiar interaction environment through messaging (N = 25) | Privacy constraints (N = 5) |
| Ease and speed of receiving information (N = 18) | |
| Development of a number of skills (N = 11) | |

## CONCLUSIONS

The purpose of this study was to examine the educational affordances and constraints of GPT-4o use by 45 students from the 2nd year of General Lyceum. The findings regarding educational affordances focus on five categories extensively analyzed in the previous section and shown in Table 1. Of these affordances, two are found in the research literature: the development of a number of skills (Prananta et al., 2023) and the ease of receiving information (Fütterer et al., 2023). The present study contributes by enriching the research literature, adding three educational affordances not previously reported.

These are: (1) the creation of new knowledge by leveraging prior knowledge, (2) immediate content feedback, and (3) the friendly interaction environment through messaging. Although the research literature references the acquisition of new knowledge with the help of AI, existing approaches focus either on the ethics of student use (Fütterer et al., 2023) or do not consider the factor of false information that GPT-4o may generate (Prananta et al., 2023). This study is one of the first to present to Greek secondary education students the constraints of content untruthfulness.

Beyond educational affordances, this study also revealed three constraints of AI use for teaching and learning. The first constraint concerns uncertainty about the truthfulness and quality of generated content. The second constraint concerned anxiety regarding AI use, and the third concerned privacy constraints. Similar findings have been reported in the research literature (e.g., Fütterer et al., 2023).

One finding of this study is that the majority of students (N = 42) before the intervention believed that the content provided by GPT-4o was truthful and that it never made errors or inaccuracies. After the intervention, students were considerably concerned about whether they could trust GPT-4o's responses. Another finding-observation concerns the difficulties students face in understanding mathematical principles. During the intervention (see Procedure), students were asked to select a subject they found difficult so that GPT-4o could adapt its explanation as if addressing younger students. Most students (N = 30) asked GPT-4o to explain the quadratic equation as if explaining it to a 10-year-old.

In conclusion, students appear to be particularly positive about using GPT-4o in teaching and learning. This study is one of the first to utilize GPT-4o with secondary education students in Greece and to examine the educational affordances and constraints of its use. Its results enrich the research literature in the field of AI for teaching and learning. Finally, it should be noted that this study used GPT-4o and students had time to complete six activities that were analyzed extensively above.

A limitation of this study concerns the sample, which was a convenience sample. However, future studies could be conducted with a larger student sample, at a different educational level, use different chatbots (e.g., DeepSeek, Gemini, Copilot, Claude), and apply this study to a teacher sample.

## [Original Greek Version / Πρωτότυπη Ελληνική Έκδοση]

*The original Greek version of this paper is appended below.*

*This paper was translated from Greek to English by the authors with the assistance of AI translation tools, in accordance with arXiv's non-English submission policy.*

# Τεχνητή Νοημοσύνη στη Δευτεροβάθμια Εκπαίδευση: Οι Εκπαιδευτικές Δυνατότητες και Περιορισμοί χρήσης του ChatGPT-4o

Τρύφων Σιβένας[1], Παναγιώτα Μαραγκάκη[2]

*[1] Μεταδιδακτορικός Ερευνητής, Εθνικό και Καποδιστριακό Πανεπιστήμιο Αθηνών, sivenastrif@primedu.uoa.gr*
*[2] Ερευνήτρια, Πανεπιστήμιο Λευκωσίας, giwmar@icloud.com*

## ΠΕΡΙΛΗΨΗ

Σκοπός της παρούσας έρευνας ήταν να εξετάσει από την πλευρά μαθητών Δευτεροβάθμιας Εκπαίδευσης, τις εκπαιδευτικές δυνατότητες και τα ζητήματα χρήσης της Τεχνητής Νοημοσύνης (ΤΝ) στη διδασκαλία και στη μάθηση. Το δείγμα αποτέλεσαν 45 μαθητές Β' τάξης Λυκείου, οι οποίοι, αφού εξοικειώθηκαν με το ChatGPT-4o και ολοκλήρωσαν έξι δραστηριότητες, συμπλήρωσαν ανοιχτού τύπου ερωτηματολόγιο σχετικά με τον σκοπό της έρευνας. Η ανοιχτή οριζόντια διαθεματική επιλεκτική κωδικοποίηση των δεδομένων (open, axial και selective coding) έδειξε ότι οι μαθητές αναγνωρίζουν τις εξής πέντε εκπαιδευτικές δυνατότητες: τη δημιουργία νέων γνώσεων βασισμένων σε παλαιότερες, την άμεση ανατροφοδότηση, τη φιλική αλληλεπίδραση μέσω μηνυμάτων, την ευκολία και τη ταχύτητα της πρόσβασης στην πληροφορία και την ανάπτυξη δεξιοτήτων. Παράλληλα, επισημάνθηκαν τρία κύρια ζητήματα: η αξιοπιστία του περιεχομένου, το άγχος για τη χρήση της ΤΝ και οι ανησυχίες για την ιδιωτικότητα. Η έρευνα καταλήγει ότι οι μαθητές είναι θετικοί απέναντι στη χρήση της ΤΝ στην εκπαίδευση.

## ΕΙΣΑΓΩΓΗ

Τα τελευταία χρόνια, οι ραγδαίες τεχνολογικές εξελίξεις στη Τεχνητή Νοημοσύνη (ΤΝ) έχουν κεντρίσει το ενδιαφέρον των ερευνητών σχετικά με την εφαρμογή και τη χρήση της σε έναν αριθμό κλάδων (Xu et al., 2021). Έναν από αυτούς αποτελεί η εκπαίδευση (Holmes & Tuomi, 2022). Πρόσφατες συστηματικές ανασκοπήσεις (Deng & Yu, 2023; Pérez et al., 2020) έχουν αναδείξει πως η ΤΝ μπορεί να συμβάλει θετικά στην αναδιαμόρφωση του τοπίου της εκπαίδευσης (Holmes & Tuomi, 2022). Μέχρι στιγμής, η έρευνα γύρω από την εφαρμογή της ΤΝ στην εκπαίδευση έχει εστιάσει τόσο σε δείγμα εκπαιδευτικών (Zawacki-Richter et al., 2019), όσο και σε δείγμα μαθητών (Deng & Yu, 2023). Σε πρόσφατη συστηματική ανασκόπησή τους, οι Zhang & Tur (2024) αναφέρουν πως η ΤΝ διευκολύνει την προσαρμοστική μάθηση, βελτιώνει και εμπλουτίζει το μαθησιακό περιβάλλον, ενώ ταυτόχρονα συμβάλει στην ανάπτυξη ενός αριθμού δεξιοτήτων των μαθητών όπως είναι η ενεργή συμμετοχή, η λήψη πρωτοβουλιών καθώς και η δημιουργικότητα.

Από τις πρόσφατες έρευνες (π.χ., Adeshola & Adepoju, 2024), ανασκοπήσεις (π.χ., Zhai et al., 2021) και συστηματικές ανασκοπήσεις (π.χ., Deng & Yu, 2023; Zhang & Tur, 2024) προκύπτει πως το εργαλείο που χρησιμοποιήθηκε κατά κύριο λόγο στις εκπαιδευτικές έρευνες για τη χρήση της ΤΝ ήταν το Chat-GPT 3.5. Το Generative Pretrained Transformer (GPT) της Open AI είναι ένα γλωσσικό μοντέλο που έχει σχεδιαστεί για να παράγει κείμενο που μοιάζει με ανθρώπινο λόγο και αποτελεί αντικείμενο ιδιαίτερου ενδιαφέροντος για την εκπαιδευτική αξιοποίησή του (Deng & Yu, 2023). Το ChatGPT βρίσκεται πλέον στην έκδοση 4o, η οποία έγινε διαθέσιμη στις 13 Μαΐου 2024 (OpenAI, 2025) και προσφέρει χωρίς πρόσθετη χρέωση έναν αυξημένο αριθμό χαρακτηριστικών και δυνατοτήτων. Χαρακτηριστικό παράδειγμα αποτελεί η δυνατότητα δημιουργίας πολυμεσικού περιεχομένου (π.χ., Εικόνας μέσω του DALL-E 2.0) και η σύνοψη επισυναπτόμενου περιεχομένου, που σε προηγούμενη έκδοση απαιτούσαν μηνιαία συνδρομή. Παρόλα αυτά, μικρός αριθμός ερευνών έχει αξιοποιήσει τη νέα έκδοση όπως επιβεβαιώνουν πρόσφατες συστηματικές ανασκοπήσεις (π.χ., Zhang & Tur, 2024).

Ειδικότερα, όπως αναφέρουν οι Baig και Yadegaridehkordi (2024), οι περισσότερες έρευνες φαίνεται να επικεντρώνονται περισσότερο στη χρήση του ChatGPT 3.5, ενώ παρατηρείται πως λιγότερες είναι εκείνες που χρησιμοποιούν την έκδοση 4.0 και τις νέες δυνατότητές της. Επιπρόσθετα, η ανασκόπηση της ερευνητικής βιβλιογραφίας επιβεβαίωσε ότι οι περισσότερες έρευνες εστιάζουν στη τριτοβάθμια εκπαίδευση (Baig & Yadegaridehkordi, 2024), ενώ η χρήση του GPT-4o στη δευτεροβάθμια εκπαίδευση δεν έχει εξεταστεί εκτεταμένα. Ένα ακόμη κενό αφορά την έλλειψη ερευνών που εξετάζουν τις δυνατότητες (affordances) που προσφέρει το GPT-4o στην εκπαίδευση.

Ο σκοπός της παρούσας έρευνας είναι να καλύψει το ερευνητικό κενό και να εξετάσει τις εκπαιδευτικές δυνατότητες (affordances) και τους περιορισμούς (constraints) της χρήσης του GPT-4o σε δείγμα 45 μαθητών της Β’ τάξης Γενικού Λυκείου. Ειδικότερα, η έρευνα απαντά στα ακόλουθα ερευνητικά ερωτήματα:

1) Ποιες είναι οι δυνατότητες του GPT-4o για τη διδασκαλία και τη μάθηση;

2) Ποια είναι τα ζητήματα που μπορούν να περιορίσουν τη χρήση του GPT-4o στη διδασκαλία και στη μάθηση;

3) Θα επέλεγαν οι μαθητές δευτεροβάθμιας εκπαίδευσης να χρησιμοποιήσουν το GPT-4o στη διδασκαλία και στη μάθηση;

## ΟΡΙΣΜΟΣ ΤΩΝ ΔΥΝΑΤΟΤΗΤΩΝ

Ο όρος δυνατότητα (affordances) αφορά την περιγραφή των προσφερόμενων δυνατοτήτων και πλεονεκτημάτων που μπορεί να παρέχει μια τεχνολογία στην εκπαίδευση (Sivenas & Koutromanos, 2022a). Δεδομένου ότι η χρήση του GPT-4o στην εκπαίδευση μπορεί να έχει ευρεία χρήση στο μέλλον, είναι σημαντικό να εξεταστούν οι δυνατότητές του στη διδασκαλία και στη μάθηση. Όπως αναφέρουν οι Masoudi et al. (2019), η εξέταση των δυνατοτήτων μιας τεχνολογίας στην εκπαίδευση βοηθά στην καλύτερη κατανόηση των χαρακτηριστικών της και του τρόπου με τον οποίο μπορούν να συμβάλουν στη διδασκαλία και στη μάθηση. Τον όρο δυνατότητα εισήγαγε ο ψυχολόγος Gibson στα τέλη της δεκαετίας του 1970 (Gibson, 1977), ενώ ο Norman (1999) εμβάθυνε στις αντιληπτές δυνατότητες (perceived affordances) στο πλαίσιο αλληλεπίδρασης ανθρώπου-υπολογιστή (HCI).

Ειδικότερα, οι αντιληπτές δυνατότητες, αφορούν αυτά που αντιλαμβάνεται ένα άτομο γύρω από ένα αντικείμενο που θα προσδιορίσουν τον τρόπο με τον οποίο θα το χρησιμοποιήσει (Norman, 1999). Ο Norman (1999) αναφέρει πως «*ο όρος δυνατότητες εστιάζει στις αντιληπτές και πραγματικές ιδιότητες που έχει ένα αντικείμενο, δίνοντας βαρύτητα στις θεμελιώδεις ιδιότητες που καθορίζουν ακριβώς τον τρόπο με τον οποίο μπορεί να χρησιμοποιηθεί ένα αντικείμενο*». Σύμφωνα με τα ανωτέρω, η παρούσα έρευνα υιοθετεί την προσέγγιση του Norman, καθώς σχετίζεται με τις ανάγκες της παρούσας έρευνας και την αλληλεπίδραση μεταξύ ανθρώπου-υπολογιστή.

## ΑΝΑΣΚΟΠΗΣΗ ΤΗΣ ΕΡΕΥΝΗΤΙΚΗΣ ΒΙΒΛΙΟΓΡΑΦΙΑΣ

Πραγματοποιήθηκε μια ανασκόπηση της ερευνητικής βιβλιογραφίας στις ακόλουθες βάσεις δεδομένων, (1) IEEEXPLORE, (2) ERIC, (3) SpringerLink, (4) Scopus καθώς και (5) Science Direct. Χρησιμοποιήθηκαν οι ακόλουθες λέξεις κλειδιά “affordances OR educational affordances OR constraints AND artificial intelligence, OR AI, OR chatbot, OR chatgpt OR openAI”.

Τα κριτήρια ήταν τρία και αφορούσαν (1) την αναζήτηση ερευνών από το 2004 και μετά, (2) έρευνες στην αγγλική και ελληνική γλώσσα και (3) έρευνες με εμπειρικά δεδομένα.

Τα αποτελέσματα της ανασκόπησης της ερευνητικής βιβλιογραφίας δεν ανέδειξαν έρευνες που να εξετάζουν τις εκπαιδευτικές δυνατότητες και ζητήματα χρήσης της ΤΝ στη διδασκαλία και στη μάθηση. Από την άλλη πλευρά, ανέδειξε έναν αριθμό ερευνών που εξετάζει τις εκπαιδευτικές δυνατότητες αντίστοιχων αναδυόμενων τεχνολογιών.

Μια από τις πρώτες έρευνες είναι εκείνη των Conole & Dyke (2004) που εξέτασαν τις δυνατότητες των ΤΠΕ και τον βαθμό που διευκολύνουν την εκπαιδευτική χρήση τους. Οι ερευνητές ανέδειξαν επτά εκπαιδευτικές δυνατότητες μεταξύ άλλων, την πρόσβαση σε μεγάλο όγκο πληροφοριών, την αμεσότητα σε ταχέως μεταβαλλόμενες πληροφορίες και την επικοινωνία και τη συνεργασία. Αντίστοιχες έρευνες γύρω από τις εκπαιδευτικές δυνατότητες πραγματοποίησαν οι Dalgarno & Lee (2010) που εξέτασαν τις εκπαιδευτικές δυνατότητες των τρισδιάστατων εικονικών μαθησιακών περιβαλλόντων, ο Song (2011) που εξέτασε τις εκπαιδευτικές δυνατότητες των φορητών PDA, οι Bower & Sturman (2015) που εξέτασαν τις εκπαιδευτικές δυνατότητες των έξυπνων φορετών τεχνολογιών και ακόμα πιο πρόσφατα η έρευνα των Sivenas και Koutromanos (2022b) που εξέτασαν τις εκπαιδευτικές δυνατότητες των drones για τη διδασκαλία και τη μάθηση.

## ΜΕΘΟΔΟΛΟΓΙΑ

Η παρούσα έρευνα υλοποιήθηκε τον Μάιο του 2024 σε τρία τμήματα Β' τάξης Γενικού Λυκείου στην Αττική. Πιο συγκεκριμένα, η παρέμβαση πραγματοποιήθηκε στο εργαστήριο του σχολείου, ώστε να έχουν οι μαθητές πρόσβαση σε υπολογιστές.

### Συμμετέχοντες

Στην έρευνα συμμετείχαν 45 μαθητές Β' τάξης Γενικού Λυκείου, εκ των οποίων 26 (57.78%) ήταν άντρες και 19 (42.22%) ήταν γυναίκες. Από τους 45, οι 18 ανέφεραν πως είχαν χρησιμοποιήσει εφαρμογή ΤΝ στο παρελθόν.

### Ερωτηματολόγιο

Η συλλογή των δεδομένων πραγματοποιήθηκε μέσω ενός ανώνυμου διαδικτυακού ερωτηματολογίου (google forms) και αποτελούνταν από δύο μέρη. Το πρώτο μέρος περιελάμβανε ερωτήσεις δημογραφικών δεδομένων (π.χ., φύλλο, ηλικία), καθώς και μια ερώτηση που ζητούσε από τους μαθητές να απαντήσουν αν είχαν χρησιμοποιήσει εφαρμογή ΤΝ στο παρελθόν. Το δεύτερο μέρος περιελάμβανε πέντε ερωτήσεις ανοιχτού τύπου και ζητούσε από τους μαθητές να απαντήσουν σχετικά με 1) τις εκπαιδευτικές δυνατότητες που εντοπίζουν στο GPT-4o, 2) το αν θα χρησιμοποιούσαν στο μέλλον GPT-4o για τη διδασκαλία και τη μάθηση, 3) τον τρόπο που φαντάζονται πως θα χρησιμοποιούσαν το GPT-4o για τη διδασκαλία και τη μάθηση καθώς και 4) σε ποιον βαθμό θα χρησιμοποιούσαν το GPT-4o για προσωπική χρήση. Η πέμπτη ερώτηση αφορούσε τα ζητήματα τα οποία πιθανόν δρούσαν ανασταλτικά στη χρήση ΤΝ στη διδασκαλία και στη μάθηση. Οι ανωτέρω ερωτήσεις παραμετροποιήθηκαν από παρόμοιες έρευνες στην ερευνητική βιβλιογραφία (π.χ., Bower & Sturman, 2015; Koutromanos et al., 2024; Sivenas & Koutromanos, 2022a).

### Εξοπλισμός

Συνολικά αξιοποιήθηκαν 20 επιτραπέζιοι υπολογιστές του εργαστηρίου πληροφορικής του σχολείου. Για την παρέμβαση δεν έγινε η εγκατάσταση της εφαρμογής ChatGPT desktop τοπικά στους υπολογιστές του εργαστηρίου, αντ' αυτού οι μαθητές χρησιμοποίησαν τον προ-εγκατεστημένο browser, ώστε να επισκεφθούν την ιστοσελίδα της Open AI.

### Διαδικασία

Η παρέμβαση πραγματοποιήθηκε σε δύο στάδια. Στο πρώτο στάδιο έγινε μια 45-λέπτη εισαγωγική παρουσίαση αναφορικά με τη ΤΝ, την ιστορική αναδρομή της, την εξέλιξή της και τις εφαρμογές της στην καθημερινότητα. Στο δεύτερο στάδιο οι εκπαιδευτικοί καθοδήγησαν τους μαθητές, ώστε να δημιουργήσουν ατομικούς λογαριασμούς στο GPT-4o και να εξοικειωθούν με το περιβάλλον του. Ακολούθως, από τους μαθητές ζητήθηκε να πραγματοποιήσουν έξι δραστηριότητες που είχαν σκοπό την εξοικείωσή τους με τις δυνατότητες του GPT-4o.

Η πρώτη δραστηριότητα ζητούσε από τους μαθητές να αναζητήσουν πληροφορίες από το GPT-4o σχετικά με ένα αντικείμενο του ενδιαφέροντος τους. Η δεύτερη δραστηριότητα αφορούσε τη δημιουργία ενός προσωποποιημένου βιογραφικού σημειώματος. Ειδικότερα, οι εκπαιδευτικοί παρουσίασαν το ακόλουθο σενάριο «*Φανταστείτε πως τώρα το καλοκαίρι που θα κλείσουν τα σχολεία* (σημ. Η έρευνα έγινε Μάιο του 2024) *θέλετε να κάνετε μια καλοκαιρινή εργασία για τον Ιούλιο και για τον Αύγουστο. Ζητήστε από το ChatGPT να φτιάξει ένα βιογραφικό σύμφωνα με τα χόμπι σας και με το αντικείμενο που θέλετε να ασχοληθείτε το καλοκαίρι*». Εν συνεχεία, οι μαθητές έδιναν προσωπικά στοιχεία (π.χ., όνομα, επώνυμο, χόμπι, πιστοποιήσεις ξένων γλωσσών, στοιχεία επικοινωνίας) και το GPT-4o δημιουργούσε ένα προσωποποιημένο βιογραφικό που οι μαθητές αποθήκευσαν ως έγγραφο στο Microsoft Word για μελλοντική χρήση. Η τρίτη δραστηριότητα αφορούσε τη χρήση του DALL-E 2.0 για τη δημιουργία εικόνων.

Η τέταρτη δραστηριότητα αφορούσε τη σύνοψη περιεχομένου. Σε αυτό το σημείο ζητήθηκε από τους μαθητές να αναζητήσουν στο διαδίκτυο κείμενα ή έγγραφα σχετικά με ένα αντικείμενο του ενδιαφέροντος τους. Έπειτα, ανέβαζαν το κείμενο ή το έγγραφο στο GPT-4o και ζητούσαν να γίνει σύνοψη περιεχομένου. Τέλος, έκαναν ερωτήσεις στο GPT-4o σχετικά με το περιεχόμενο που είχαν ανεβάσει. Για την πέμπτη δραστηριότητα ζητήθηκε από τους μαθητές να δημιουργήσουν με το GPT-4o ένα διαγώνισμα που αποτελείται από πέντε ερωτήσεις για ένα μάθημα του σχολείου που τους δυσκολεύει.

Ακολούθως, οι μαθητές ζήτησαν από το GPT-4o να απαντήσει τις ερωτήσεις που δημιούργησε. Η δραστηριότητα αυτή είχε σκοπό να αναδείξει στους μαθητές πως το GPT-4o μπορεί να απαντήσει λανθασμένα. Η έκτη και τελευταία δραστηριότητα, ζητούσε από τους μαθητές να αξιοποιήσουν το GPT-4o και να ζητήσουν πληροφόρηση και επεξήγηση δύσκολων όρων με το GPT-4o να απευθύνεται σε διαφορετική ηλικιακή ομάδα. Ειδικότερα, οι εκπαιδευτικοί ζήτησαν από τους μαθητές να επιλέξουν ένα μάθημα που αντιμετωπίζουν δυσκολίες (π.χ., μαθηματικά). Έπειτα, τους ζήτησαν να επιλέξουν ένα συγκεκριμένο αντικείμενο του μαθήματος που τους δυσκολεύει (π.χ., δευτεροβάθμια εξίσωση). Από τους μαθητές ζητήθηκε να γράψουν στο GPT-4o «Εξήγησε μου το Χ σαν να απευθύνεσαι σε παιδί 7, 10 και 14 ετών». Η διάρκεια του δεύτερου σταδίου ήταν 45 λεπτά.

Εν συνεχεία, οι μαθητές είχαν 15 λεπτά στη διάθεσή τους, ώστε να κάνουν ελεύθερη χρήση του GPT-4o. Τέλος, πραγματοποιήθηκε η συμπλήρωση του διαδικτυακού ερωτηματολογίου, η οποία είχε διάρκεια δεκαπέντε λεπτά.

### Ανάλυση των δεδομένων

Η ανάλυση των δεδομένων έγινε από δύο ειδικούς σε θέματα ΤΠΕ στην εκπαίδευση και βασίστηκε στις αρχές της ποιοτικής έρευνας και θεματικής ανάλυσης των Corbin & Strauss (2014). Ειδικότερα, έγινε ανοιχτή οριζόντια διαθεματική επιλεκτική κωδικοποίηση των δεδομένων (open, axial και selective coding) (Corbin & Strauss, 2014). Παρόμοια μεθοδολογική προσέγγιση έχουν ακολουθήσει προηγούμενες έρευνες στην ερευνητική βιβλιογραφία που εξέτασαν τις εκπαιδευτικές δυνατότητες τεχνολογιών (π.χ., Bower & Sturman, 2015; Koutromanos et al., 2024; Sivenas & Koutromanos, 2022b).

## ΑΠΟΤΕΛΕΣΜΑΤΑ

Τα αποτελέσματα ανέδειξαν πέντε εκπαιδευτικές δυνατότητες της ΤΝ όπως φαίνονται στον Πίνακα 1. Συνολικά, η πλειοψηφία των συμμετεχόντων (N=42) αναγνώρισε τα οφέλη που μπορεί να προσφέρει στην εκπαίδευση. Παρακάτω ακολουθεί η ανάλυση των εκπαιδευτικών δυνατοτήτων σε ξεχωριστή υποκατηγορία.

**Δημιουργία νέων γνώσεων αξιοποιώντας παρελθούσες γνώσεις**
Η πλειοψηφία των συμμετεχόντων (N=33) αναγνώρισε τη δυνατότητα απόκτησης νέων γνώσεων ως μια ιδιαίτερα σημαντική εκπαιδευτική δυνατότητα. Ειδικότερα, οι μαθητές έδωσαν έμφαση στο γεγονός πως θα χρησιμοποιούσαν το GPT-4o για να συζητήσουν σχετικά με θέματα που ήδη γνώριζαν ή σχετικά με θέματα που γνωρίζουν εν μέρη κάποια στοιχεία. Αυτό οφείλεται στο γεγονός πως κατανοήσαν κατά τη διάρκεια της παρέμβασης πως το GPT-4o μπορεί να κάνει λάθη. Από την άλλη πλευρά, ένας αριθμός μαθητών (N=13) ανέφερε πως θα χρησιμοποιούσε το GPT-4o για δημιουργία νέων γνώσεων χωρίς να λαμβάνουν υπόψιν τους την πιθανότητα σφάλματος.

**Άμεση ανατροφοδότηση περιεχομένου**
Μια εκπαιδευτική δυνατότητα που ανέδειξε μεγάλη μερίδα των μαθητών (N=28) αφορά την άμεση ανατροφοδότηση που προσφέρει το GPT-4o. Ειδικότερα, εντοπίστηκαν δύο κατηγορίες απαντήσεων. Η πρώτη αφορά το γεγονός πως η άμεση ανατροφοδότηση αφορά σε περιεχόμενο που έγραψαν οι ίδιοι οι μαθητές στο GPT-4o, δηλαδή το στάδιο της συνομιλίας. Έτσι, αναφέρουν πως το GPT-4o τους βοήθησε, ώστε να γίνει καλύτερο το κείμενο τους ή να είναι σε θέση να περιγράφει στοχευμένα αυτό που οι μαθητές προσπαθούν να γράψουν. Η δεύτερη κατηγορία απαντήσεων αφορά τη σύνοψη, δηλαδή την παρέμβαση που έκαναν οι μαθητές όπου ανέβασαν ένα συνημμένο αρχείο και ζήτησαν να γίνει η σύνοψη περιεχομένου. Χαρακτηριστικό παράδειγμα αποτελεί η απάντηση του μαθητή No14 που ανέφερε πως: «*ανέβασα ένα pdf με ένα βιβλίο για το μεταπτυχιακό της αδερφής μου που είχα στο φλασάκι* (εννοεί usb thumb drive). *Το βιβλίο είναι 500 σελίδες, αλλά με το AI έκανε τη σύνοψη σε δύο λεπτά, είναι πολύ γρήγορο δεν ξέρω αν θα είχα υπομονή να περιμένω μια ώρα για να το κάνει πάντως*».

**Φιλικό/οικείο περιβάλλον αλληλεπίδρασης μέσω μηνυμάτων**
Οι μαθητές αναφέρθηκαν στο φιλικό περιβάλλον χρήσης μέσω αποστολής μηνυμάτων (N=25). Όπως ανέφεραν αυτός ο τρόπος επικοινωνίας είναι πολύ γνώριμος και φιλικός και είναι ο τρόπος επικοινωνίας που επιλέγουν για συνομιλία με φίλους. Αυτό είχε ως αποτέλεσμα η συνομιλία με το GPT-4o να τους φαντάζει πιο οικεία.

**Ευκολία και ταχύτητα στη λήψη της πληροφορίας**
Άλλη μια δυνατότητα που ανέδειξαν οι μαθητές αφορά την ταχύτητα με την οποία λαμβάνουν απαντήσεις από το GPT-4o (N=18). Ειδικότερα, ο χρόνος ανταπόκρισης είναι ιδιαίτερα σημαντικός για τους μαθητές, με αρκετούς να σχολιάζουν πως σε περίπτωση αργής ανταπόκρισης ίσως να μην το χρησιμοποιούσαν.

**Ανάπτυξη ενός αριθμού δεξιοτήτων**
Αρκετοί μαθητές αναφέρθηκαν έμμεσα στη συμβολή του GPT-4o για την ανάδειξη δεξιοτήτων (N=11). Αναλυτικότερα, όπως ανέφεραν οι μαθητές πιστεύουν πως το GPT-4o μπορεί να τους βοηθήσει να βελτιωθούν σε ορισμένα μαθήματα, με αποτέλεσμα να έχουν ενεργή συμμετοχή στη διδασκαλία. Άλλα παραδείγματα μαθητών εστίασαν στην κατανόηση του μαθήματος της πληροφορικής και συγκεκριμένα εντολών προγραμματισμού. Οι υπόλοιπες δεξιότητες που ανέφεραν αφορούσαν τον ψηφιακό γραμματισμό (N=8), την υπολογιστική σκέψη (N=6), την επίλυση προβλημάτων (N=4), την κριτική σκέψη (N=3) και την επικοινωνία (N=2).
Σχετικά με την τελευταία ερευνητική ερώτηση που εξέταζε τα ζητήματα και τους πιθανούς περιορισμούς που μπορεί να επηρεάσουν τη χρήση του GPT-4o στη διδασκαλία και στη μάθηση εντοπίστηκαν τρείς κατηγορίες (βλ. Πίνακα 1) που αναλύονται παρακάτω.

**Άγνοια σχετικά με την αλήθεια και την ποιότητα του περιεχομένου**
Το πρώτο ζήτημα που ανέδειξαν αρκετοί μαθητές αφορούσε την άγνοια/αβεβαιότητα αναφορικά με την αλήθεια και την ποιότητα του περιεχομένου (N=16). Ειδικότερα, αρκετοί μαθητές αναφέρθηκαν στο γεγονός πως μετά την παρέμβαση δεν θα επέλεγαν το GPT-4o για να αποκτήσουν νέες γνώσεις, παρά μόνο θα ζητούσαν επιπρόσθετες πληροφορίες στις γνώσεις που ήδη έχουν. Από την άλλη πλευρά, κάποιοι μαθητές αναφέρθηκαν στο γεγονός πως δρα ως ανασταλτικός παράγοντας, αφού μετά από κάθε αναζήτηση πληροφοριών στο GPT-4o θα χρειάζονται επιπρόσθετο χρόνο για να αναζητούν στο διαδίκτυο το αν τα αποτελέσματα/δεδομένα που προσέφερε το GPT-4o είναι αληθή.

**Άγχος αναφορικά με τη χρήση της ΤΝ**
Ένας αριθμός μαθητών ανέφερε πως η χρήση του GPT-4o τους δημιουργεί άγχος και στρες (N=8). Οι περισσότερες απαντήσεις βασίστηκαν στη δυνατότητα ανατροφοδότησης περιεχομένου. Τα περισσότερα σχόλια των μαθητών αφορούσαν το γεγονός πως το κείμενο που εκείνοι έγραψαν/δημιουργούσαν δεν θα ικανοποιούσε ποτέ το GPT-4o. Χαρακτηριστικό είναι το παράδειγμα του μαθητή Νο 32 που ανέφερε στην απάντησή του: «*Με αγχώνει που κάθε φορά που ρωτάω πως μπορώ να βελτιώσω το κείμενο που έγραψα, συνέχεια μου κάνει σχόλια. Δηλαδή το κείμενο μου ποτέ δεν είναι τέλειο; Συνέχεια μου προτείνει αλλαγές και πως θα γίνει καλύτερο, αυτό με στρεσάρει κάποια στιγμή πρέπει να σταματήσει και να πει είναι οκ*». Μικρότερος αριθμός μαθητών αναφέρθηκε στο γεγονός πως νιώθουν άγχος επειδή όπως αναφέρει χαρακτηριστικά η μαθήτρια Νο 9: «*…μπορώ να μιλήσω με μια οντότητα που ξέρει τα πάντα και αυτό είναι αγχωτικό*».

**Ζητήματα ιδιωτικότητας (N=5)**
Λίγοι μαθητές αναφέρθηκαν στα ζητήματα ιδιωτικότητας του GPT-4o (N=5). Οι περισσότερες απαντήσεις αφορούσαν προβληματισμούς των μαθητών αναφορικά με το αν κάποιος μπορεί να δει τις πληροφορίες που ρωτάνε στο GPT-4o, καθώς και το που αυτές αποθηκεύονται. Άλλες απαντήσεις αφορούσαν την προστασία της ιδιωτικότητας τους, όπου οι μαθητές δεν επιθυμούσαν να δηλώσουν το όνομα τους κατά την εγγραφή στην σελίδα της open AI.

**Πίνακας 1**
Εκπαιδευτικές δυνατότητες και ζητήματα χρήσης της Τεχνητής Νοημοσύνης στη Δευτεροβάθμια Εκπαίδευση

| **Δυνατότητες του GPT-4o για τη διδασκαλία και τη μάθηση** | **Ζητήματα από τη χρήση** |
|---|---|
| Δημιουργία νέων γνώσεων αξιοποιώντας παρελθούσες γνώσεις (N=33) | Άγνοια σχετικά με την αλήθεια και την ποιότητα του περιεχομένου (N=16) |
| Άμεση ανατροφοδότηση περιεχομένου (N=28) | Άγχος αναφορικά με τη χρήση ΤΝ (N=8) |
| Φιλικό/οικείο περιβάλλον αλληλεπίδρασης μέσω μηνυμάτων (N=25) | Ζητήματα ιδιωτικότητας (N=5) |
| Ευκολία και ταχύτητα στη λήψη της πληροφορίας (N=18) | |
| Ανάπτυξη ενός αριθμού δεξιοτήτων (N=11) | |

**ΣΥΜΠΕΡΑΣΜΑΤΑ**

Σκοπός της παρούσας έρευνας ήταν η εξέταση των εκπαιδευτικών δυνατοτήτων, καθώς και των ζητημάτων χρήσης του GPT-4o από 45 μαθητές Β' τάξης Γενικού Λυκείου. Τα ευρήματα όσον αφορά τις εκπαιδευτικές δυνατότητες εστιάζουν σε πέντε κατηγορίες που αναλύθηκαν εκτεταμένα στην προηγούμενη ενότητα και φαίνονται στον Πίνακα 1. Από τις ανωτέρω δυνατότητες δύο εντοπίζονται στην ερευνητική βιβλιογραφία, πρόκειται για την ανάπτυξη ενός αριθμού δεξιοτήτων (Prananta et al., 2023), καθώς και την ευκολία στη λήψη της πληροφορίας (Fütterer et al., 2023). Η παρούσα έρευνα συμβάλει εμπλουτίζοντας την ερευνητική βιβλιογραφία, προσθέτοντας τρεις εκπαιδευτικές δυνατότητες που δεν έχουν αναφερθεί μέχρι στιγμής στην ερευνητική βιβλιογραφία.

Πρόκειται για (1) την δημιουργία νέων γνώσεων αξιοποιώντας παρελθούσες γνώσεις, (2) την άμεση ανατροφοδότηση περιεχομένου καθώς και (3) το φιλικό περιβάλλον αλληλεπίδρασης μέσω μηνυμάτων. Παρόλο που στην ερευνητική βιβλιογραφία γίνεται αναφορά στην απόκτηση νέων γνώσεων με τη βοήθεια της ΤΝ, οι μέχρι τώρα προσεγγίσεις εστιάζουν είτε στην ηθικότητα της χρήσης του από μαθητές (Fütterer et al., 2023), είτε δεν λαμβάνουν υπόψιν τους τον παράγοντα της αναληθούς πληροφορίας που μπορεί να δημιουργήσει το GPT-4o (Prananta et al., 2023). Η παρούσα έρευνα αποτελεί μια από τις πρώτες που παρουσίασε σε μαθητές Δευτεροβάθμιας Ελληνικής Εκπαίδευσης τα ζητήματα της αναλήθειας περιεχομένου.

Πέραν των εκπαιδευτικών δυνατοτήτων, η παρούσα έρευνα ανέδειξε και τρία ζητήματα χρήσης της ΤΝ για τη διδασκαλία και τη μάθηση. Το πρώτο ζήτημα αφορά την άγνοια σχετικά με την αλήθεια και την ποιότητα του παραγόμενου περιεχομένου. Το δεύτερο αφορούσε άγχος αναφορικά με τη χρήση της ΤΝ και το τρίτο αφορούσε ζητήματα ιδιωτικότητας. Παρόμοιά ευρήματα έχουν αναδείξει έρευνες στην ερευνητική βιβλιογραφία (π.χ., Fütterer et al., 2023).

Ένα εύρημα της παρούσας έρευνας είναι πως η πλειοψηφία των μαθητών (N=42) πριν την υλοποίηση της παρέμβασης θεωρούσε πως το περιεχόμενο που προσφέρει το GPT-4o είναι αληθές και πως σε καμία περίπτωση δεν κάνει λάθη και ανακρίβειες. Μετά την παρέμβαση οι μαθητές προβληματίστηκαν αρκετά σχετικά με το αν είναι πλέον σε θέση να εμπιστευτούν τις απαντήσεις του GPT-4o. Ένα ακόμη εύρημα - παρατήρηση έχει να κάνει με τις δυσκολίες που αντιμετωπίζουν οι μαθητές στην κατανόηση αρχών των μαθηματικών. Κατά τη διάρκεια της παρέμβασης (βλ. Διαδικασία) ζητήθηκε από τους μαθητές να επιλέξουν ένα μάθημα που τους δυσκολεύει, ώστε το GPT-4o να προσαρμόσει την επεξήγηση του σαν να απευθύνεται σε μαθητές μικρότερης ηλικίας. Οι περισσότεροι μαθητές (N=30) ζήτησαν από το GPT-4o να τους εξηγήσει την δευτεροβάθμια εξίσωση σαν να την εξηγεί σε παιδί 10 χρονών.

Συμπερασματικά, οι μαθητές φαίνεται να είναι ιδιαίτερα θετικοί σχετικά με τη χρήση του GPT-4o στη διδασκαλία και στη μάθηση. Η έρευνα αυτή αποτελεί μία από τις πρώτες που αξιοποίησε το GPT-4o σε μαθητές Δευτεροβάθμιας Εκπαίδευσης στην Ελλάδα και εξέτασε τις εκπαιδευτικές δυνατότητες και τα ζητήματα χρήσης του. Τα αποτελέσματά της εμπλουτίζουν την ερευνητική βιβλιογραφία στο πεδίο της ΤΝ για τη διδασκαλία και τη μάθηση. Τέλος, πρέπει να τονιστεί πως στην παρούσα έρευνα χρησιμοποιήθηκε το GPT-4o και οι μαθητές είχαν στη διάθεση τους χρόνο ώστε να πραγματοποιήσουν έξι δραστηριότητες που αναλύθηκαν εκτενώς ανωτέρω.

Ένας περιορισμός της παρούσας έρευνας αφορά το δείγμα, το οποίο ήταν βολικό. Ωστόσο, μελλοντικές έρευνες θα μπορούσαν να υλοποιηθούν σε μεγαλύτερο δείγμα μαθητών, σε διαφορετική βαθμίδα εκπαίδευσης, να χρησιμοποιήσουν διαφορετικά chatbots (π.χ., DeepSeek, Gemini, Copilot, Claude) και να εφαρμόσουν την παρούσα έρευνα σε δείγμα εκπαιδευτικών.